\documentclass[aps,twocolumn,groupedaddress,amsmath,amssymb]{revtex4}%
\usepackage{amsmath, dsfont}
\usepackage{amssymb}
\usepackage[dvips]{graphics}
\usepackage{epsfig}
\usepackage{calc}
\usepackage{amsfonts}
\usepackage{graphicx}
\usepackage[ansinew]{inputenc}%
\usepackage{amsmath}%
\providecommand{\U}[1]{\protect\rule{.1in}{.1in}}

\begin{document}
\title[ ]{Black holes in New Massive Gravity dressed by a (non)minimally coupled scalar field}
\author{Francisco Correa}
\email{correa-at-cecs.cl}
\affiliation{Centro de Estudios Científicos (CECs), Valdivia, Chile.}
\author{Mokhtar Hassaine}
\email{hassaine-at-inst-mat.utalca.cl}
\affiliation{Instituto de Matem\'atica y Física, Universidad de Talca, Casilla 747, Talca, Chile.}
\author{Julio Oliva}
\email{julio.oliva-at-uach.cl.}
\affiliation{Instituto de Ciencias Físicas y Matem\'aticas, Universidad Austral de Chile,
Valdivia, Chile}

\begin{abstract}
We consider a self-interacting, massive scalar field (non)minimally coupled
to new massive gravity in three dimensions. For this model, we first derive a
family of black hole solutions depending on a unique integration constant and
parameterized in terms of the nonminimal coupling parameter. Imposing the absence of naked singularities restricts the parameters in
such a way that the field vanishes at infinity and fixes the metric to be
asymptotically AdS. Within this family of solutions it is possible to find a
black hole supported by a minimally coupled scalar field and therefore the
existence of these solutions is not inherent to the presence of the nonminimal
coupling. The Wald formula for
the entropy, being proportional to the lapse function evaluated at the
horizon, yields a zero entropy in spite of the fact that the solution has a
non zero temperature. As a consequence, the unique integration constant may be
interpreted as a sort of gravitational hair. As in the source free case, we
show that the same field equations admit as well asymptotically Lifshitz black
hole solutions in a different region of the space of parameters. These Lifshitz
solutions are divided in three families for which all the parameters entering
in the action may be expressed in term of the dynamical exponent.

\end{abstract}
\maketitle

\setcounter{page}{1}


\section{Introduction}

From the discovery of the BTZ black hole solution \cite{Banados:1992wn}, there
has been an increasing interest for black hole physics in three dimensions.
This interest has taken much bigger proportion the last two decades in part
because of their implication in the context of AdS$_{3}$/CFT$_{2}$
correspondence. Indeed, it is now well accepted that three-dimensional gravity
is an excellent laboratory in order to explore and test some of the ideas behind the AdS/CFT
correspondence \cite{Maldacena:1997re}. In contrast with the
four-dimensional case, the existence of the BTZ black hole in three dimensions
is inherent to the presence of the negative cosmological constant and due to
the lack of local degrees of freedom in pure Einstein gravity in three dimensions
it is the global structure of this solution what provides it's non-triviality. A way to circumvent this behavior was proposed long time ago in
\cite{Deser:1981wh} by adding to the standard Einstein-Hilbert action a
Chern-Simons term built out of the connection. In second order formulation, the resulting theory called topologically massive gravity which is
not parity invariant because of the Chern-Simons term, propagates a single
massive spin 2 (left or right) degree of freedom depending on the sign of the
Chern-Simons coupling parameter. More recently, a massive gravity respecting
parity invariance has been proposed in Ref. \cite{Bergshoeff:2009hq}, where
the authors showed that a particular and specific combination of the quadratic
invariants constructed with the scalar curvature $R$ and the Ricci tensor
$R_{\mu\nu}$ kill the dangerous spin 0 mode leaving only a healthy massive spin 2 field.
Soon after its presentation, this theory whose action in presence of
cosmological term $\Lambda$ is given by
\small{
\begin{equation}\notag
S_{NMG}=\frac{1}{2}\int  d^{3}x\sqrt{-g}\, \left[  R-2\Lambda-\frac{1}{m^{2}}
\left(  R_{\mu\nu}R^{\mu\nu}-\frac{3}{8}R^{2}\right)  \right] ,
\end{equation}}
was dubbed new massive gravity. This new gravity theory has been intensively
studied during the last years, and shown to admit interesting physical
solutions as for examples AdS waves \cite{AyonBeato:2009yq}, warped AdS black
hole \cite{Clement:2009gq}, type $D$ and type $N$ solutions, see
\cite{Ahmedov:2010uk} and \cite{Ahmedov:2011yd}. It is also interesting to
stress that new massive gravity accommodates black hole solutions with rather
different asymptotic behaviors. Indeed, within the black hole spectra of
solutions of the theory, one finds rotating asymptotically AdS black holes
with a gravitational hair \cite{Oliva:2009ip} as well as a Lifshitz black hole
with a dynamical exponent $z=3$ \cite{AyonBeato:2009nh}. By Lifshitz black
hole, we mean a black hole spacetime whose asymptotic behaviour enjoys of an
anisotropic scaling symmetry where time and space scale with different
weights. Recently, there has been an intense activity to promote the ideas
underlying the gauge-gravity duality to non-relativistic physics with the hope
of gaining a better understanding of some strongly coupled condensed matter
physics phenomena observed in laboratories, for a review see e. g.
\cite{Hartnoll:2009sz,Herzog:2009xv}. In this context, Lifshitz spacetimes
whose metrics read
\begin{equation}
ds_{\mathcal{L}}^{2}=-{r^{2z}}dt^{2}+\frac{dr^{2}}{r^{2}} +{r^{2}}d\vec
{x}_{D-2}^{2} \,,\label{Lifshitz}%
\end{equation}
are natural candidates to be the gravity duals for non-relativistic scale
invariant theories, \cite{Kachru:2008yh}. Here, the dynamical exponent $z$
reflects the anisotropy of the scaling symmetry
\[
t\to\lambda^{z} t,\qquad r\to\frac{r}{\lambda},\qquad\vec{x}\to\lambda\vec
{x} \,.
\]

In the present work, we establish that new massive gravity in three dimensions
may also accommodate black hole solutions with a source given by a
(non)minimally coupled and self-interacting scalar field whose action reads
{\small
\begin{align}\notag
S_{M}=-\int d^{3}x\sqrt{-g}\left[ \frac{1}{2} \partial_{\mu}\Phi\partial^{\mu
}\Phi+\frac{\xi}{2}R\Phi^{2}+\frac{M}{2}\Phi^{2}+\frac{\lambda}{4!}\Phi
^{4}\right] \, .\label{sfnc}%
\end{align}
} Here, $\xi\geq0$ denotes the nonminimal coupling parameter, $R$ the scalar
curvature, $M$ is a constant identified as part of the mass of the scalar
field and $\lambda$ is the coupling constant of the potential $\Phi^{4}$. The
field equations obtained by varying the action $S_{NMG}+S_{M}$ read
\begin{subequations}
\label{eqsmotionb}%
\begin{align}
& G_{\mu\nu}+\Lambda{g}_{\mu\nu}-\frac{1}{2m^{2}}K_{\mu\nu}=T_{\mu\nu}\,,\\
& \Box\Phi= \xi R \Phi+M\Phi+\frac{\lambda}{3!}\Phi^{3}\,,
\end{align}
where we have defined
\end{subequations}
\begin{align}
K_{\mu\nu}  & =2\square{R}_{\mu\nu}-\frac{1}{2}\nabla_{\mu} \nabla_{\nu}%
{R}-\frac{1}{2}\square{R}g_{\mu\nu} +4R_{\mu\alpha\nu\beta}R^{\alpha\beta
}\\ &  -\frac{3}{2}RR_{\mu\nu}-R_{\alpha\beta}R^{\alpha\beta}g_{\mu\nu} +\frac{3}{8}R^{2}g_{\mu\nu}\,.
\label{eom}
\end{align}
and the stress tensor is given by
\begin{align}
T_{\mu\nu}= & \partial_{\mu}\Phi\partial_{\nu} \Phi-g_{\mu\nu}\Big(\frac{1}%
{2}\,\partial_{\sigma}\Phi\partial^{\sigma}\Phi+\frac{M}{2}\Phi^{2}%
+\frac{\lambda}{4!}\Phi^{4}\Big)\nonumber\\
& +\xi\left( g_{\mu\nu}\Box-\nabla_{\mu}\nabla_{\nu}+G_{\mu\nu}\right)
\Phi^{2}.
\end{align}
In the case of pure Einstein gravity with a cosmological constant, there
exists a broad literature of black hole solutions with a (non)minimal scalar
field which started with the pioneering work of Martinez and Zanelli
\cite{Martinez:1996gn}, see e. g. \cite{Henneaux:2002wm,Correa:2010hf,Correa:2012rc,Xu:2013nia}. In this note, we revisit this problem in the
context of new massive gravity. Interestingly enough, we will show that, as in
the free source case, there exist a family of asymptotically AdS black hole
solutions as well as Lifshitz black hole configurations.

The plan of the paper is the following. In the next section, we present a
family of AdS black hole solution depending on a unique integration constant.
For this class of solutions, the constants $m$, $M$, $\Lambda$ and $\lambda$
are all parametrized in terms of the nonminimal coupling parameter and the AdS
radius $l$. We also show that these solutions may even exist for a scalar
which is minimally coupled to gravity ($\xi=0$), while if $\xi$ takes the value that
provides a conformally coupled scalar ($\xi=1/8$), the solution reduces to a
stealth black hole configuration on a particular case of the asymptotically
AdS black hole of \cite{Oliva:2009ip}-\cite{Bergshoeff:2009aq}. In Sec. $3$, three classes of
asymptotically Lifshitz black holes are reported for which all the parameters
entering in the action are fixed in terms of the dynamical exponent. The last
section is devoted to the discussion and some conclusions.


\section{Asymptotically AdS black hole solutions}

A black hole solution of the field equations (\ref{eqsmotionb}) is given by
\begin{subequations}
\begin{align}
 ds^{2}&=-F(r)dt^{2}+\frac{dr^{2}}{F(r)}+r^{2}d\varphi^{2}\, ,\label{sf}\\
 F(r)&=\frac{r^{2}}{l^{2}}-c_{1}\left(  \frac{r}{l}\right)  ^{\frac{32\xi
-5}{16\xi-3}}\, , \\
 \Phi(r)&=\left(  \frac{l}{r}\right)  ^{\frac{1}{6-32\xi}}\sqrt{\frac
{8c_{1}(32\xi-5)}{256\xi^{2}-32\xi-1}}\, ,\label{field}%
\end{align}
where $c_{1}$ is an arbitrary integration constant, and the cosmological term
in the action $\Lambda$, the self-interaction coupling $\lambda$, the graviton
mass and the scalar field mass are fixed as
\end{subequations}
\begin{align}
 \Lambda&=-\frac{(16\xi-1)(48\xi-7)}{2\left(  256\xi^{2}-32\xi-1\right)
l^{2}},\label{cc}\\
 \lambda&=-\frac{3(8\xi-1)\left(  256\xi^{2}-32\xi-1\right)  \left(
768\xi^{2}-152\xi+9\right)  }{16(16\xi-3)^{2}(32\xi-5)l^{2}}\, ,\nonumber\\
 m^{2}&=\frac{256\xi^{2}-32\xi-1}{2(16\xi-3)^{2}l^{2}}\, ,\nonumber\\
 M&=\frac{(8\xi-1)\left(  768\xi^{2}-192\xi+11\right)  }{4(16\xi-3)^{2}l^{2}%
}\, .\nonumber
\end{align}
The isolated points defined by $\xi=\frac{1}{16}(1+\sqrt{2})$,
$\xi=\frac{5}{32}$ and $\xi=\frac{3}{16}$ are excluded from this family.

Imposing the absence of naked singularities at infinity implies that the
coupling $\xi$ must be restricted such that the leading term in the lapse
should be $r^{2}$ (therefore fixing an asymptotically AdS behaviour). As a
consequence, the scalar field vanishes at infinity. Requiring in addition the
existence of an event horizon, finally implies that the range of physically
allowed values of $\xi$ is
\begin{equation}
\xi\in\lbrack0,\frac{1}{16}(1+\sqrt{2})[\,\cup\,]\frac{5}{32},\frac{3}%
{16}[\, ,\label{range2}%
\end{equation}
Clearly the strenght of the subleading term in the metric strongly depends on
the value of the nonminimal coupling parameter $\xi$.

The horizon is located at $r=r_{+}=l\,c_{1}^{3-16\xi}$ while the Hawking
temperature of these solutions is given by
\begin{equation}
T_{H}=\frac{\,c_{1}^{3-16\xi}}{4\pi l(3-16\xi)}\, .
\end{equation}
The Wald formula for the entropy $\mathcal{S}$,
being proportional to the lapse metric function evaluated at the horizon
yields a zero entropy
\begin{equation}
\mathcal{S}\propto\left( 1-c_{1}\left(\frac{r}{l}\right)^{\frac{1}{16\xi-3}}\right)  |_{r=r_{+}%
}=0\, .
\end{equation}
Note that zero entropy black hole solutions with planar horizon have also been
found for scalar fields nonminimally coupled with the general Lovelock gravity
in arbitrary dimension in \cite{Correa:2013bza}. Assuming that the first law
of thermodynamics holds, implies that the solutions in our family have zero
mass. As a consequence, one may interpret the unique integration constant
$c_{1}$ as a gravitational hair.

To conclude this section, let us analyze in more detail the solutions obtained
for some relevant particular values of $\xi$. It is remarkable to note that we are allowed
to consider the minimally coupled case $\left(  \xi=0\right)  $,
and therefore conclude that the existence of these solutions is not inherent
to the presence of the nonminimal coupling $R\Phi^{2}$ in the action as it is
the case for example in four dimensions,
\cite{Bekenstein:1974sf,Bocharova:1970,Martinez:2002ru}. For this case the
metric reduces to%
\begin{equation}
ds^{2}=-\left(  \frac{r^{2}}{l^{2}}-c_{1}\frac{r^{5/3}}{l^{5/3}}\right)
dt^{2}+\frac{dr^{2}}{\frac{r^{2}}{l^{2}}-c_{1}\frac{r^{5/3}}{l^{5/3}}}%
+r^{2}d\phi^{2}\, ,
\end{equation}
and the values for the couplings can be read from equation (\ref{cc}) while the expression for the field
is obtained from (\ref{field}) by fixing $\xi=0$.

When $\xi=\frac{1}{8}$, the scalar field becomes massless $M=0$ and not
self-interacting $\lambda=0$, giving indeed a conformal invariant matter
source. The remaining parameters take the values $\Lambda=m^{2}=-\frac{1}{2}$
while the metric reduces to%
\begin{equation}
ds^{2}=-\left(  \frac{r^{2}}{l^{2}}-c_{1}\frac{r}{l}\right)  dt^{2}%
+\frac{dr^{2}}{\frac{r^{2}}{l^{2}}-c_{1}\frac{r}{l}}+r^{2}d\phi^{2}%
\, .\label{specialtempo}%
\end{equation}
The expresion for the field can be read as well from (\ref{field}) by setting $\xi=\frac{1}{8}$.
This solution is a particular case of the asymptotically AdS black hole found
in the absence of sources for new massive gravity \cite{Oliva:2009ip,Bergshoeff:2009aq}. In other words, for the conformal coupling
$\xi=\frac{1}{8}$, the black hole becomes a particular solution of the field
equations (\ref{eqsmotionb}) for which both sides (the gravity and the source
parts) vanish identically
\[
G_{\mu\nu}+\Lambda{g}_{\mu\nu}-\frac{1}{2m^{2}}K_{\mu\nu}=0=T_{\mu\nu} \, .
\]
This kind of black hole configurations were discussed in
\cite{AyonBeato:2004ig} and dubbed stealth. Note that this
stealth solution corresponds to a particular case of the one
found in \cite{Hassaine:2013cma}.

We now show that the same field equations (\ref{eqsmotionb}) may also
accommodate Lifshitz black hole solutions for a different region of the
parameters as it occurs in the source free case with the AdS solution
\cite{Oliva:2009ip} and the $z=3$ Lifshitz solution \cite{AyonBeato:2009nh}.


\section{Asymptotically Lifshitz black hole solutions}

There are three branches of solutions with Lifshitz asymptotic for the above
equations of motion (\ref{eqsmotionb}). In all of them the solutions have the
following generic form
\begin{align}
  ds^{2}&=r^{2z}\left( 1-\frac{c_1}{r^{\chi}}\right) dt^{2}+\frac{dr^{2}%
}{\displaystyle r^{2} \left( 1-\frac{c_1}{r^{\chi}}\right) }+r^{2}
d\varphi^{2}\, ,\nonumber\\
  \Phi(r)&=\frac{\sqrt{\alpha}}{r^{\chi/2}}\, ,%
\end{align}
with the cosmological term and the graviton mass parameterized in term of the
dynamical exponent as
\begin{align}
\Lambda=-\frac{1}{2} \left( z^{2}+z+1\right) ,\qquad m^{2}=-\frac{1}{2}
(z^2-3z +1)\, ,
\end{align}
where from now on we set $l=1$.
The different branches share basically the same features than those discussed
in the AdS case, and are presented with some detail below.

\subsection{$\chi=(z+1)$}

For $\chi=(z+1)$, there exists a Lifshitz black hole solution where the
parameters are fixed as follows
\begin{align}
 \xi&=\frac{5}{32},\quad\alpha=\frac{16 c_{1}(1-z)}{z^2-3z+1},\nonumber\\
 M&=\frac{1}{16} (z^2-3z+1),\nonumber\\
 \lambda&=\frac{3(z^2-3z+1) (8z^2+11z+13)}{256(z-1)}\, .%
\end{align}
In order to deal with a real scalar field, the constant $c_{1}$ must be
strictly positive and the existence of a horizon is ensured for
\[
z\in[0, \frac{3-\sqrt{5}}{2}[\cup]1,\frac{3+\sqrt{5}}{2}[ \, .
\]
As in the previous sections the other values at the boundaries of the intervals are excluded from this family of solutions.


\subsection{$\chi=2(z-1)$}

The second family of Lifshitz black hole solutions is only valid for $z\geq1$
in order to have the correct Lifshitz asymptotic, and the parameters take the
following form
\begin{align}
 \xi&=\frac{2z-1}{4( 3z-1)} ,\quad\alpha=\frac{4 c_1 (1-3 z)}{z^2-3 z+1}\, ,\nonumber\\
  M&=\frac{(2 z-5) (z^2-3 z+1)}{6
z-2}\, ,\nonumber\\
 \lambda&=\frac{3(z^2-3 z+1) (12z^3-44z^2+48z-13) }{4 (1-3 z)^{2}}\, .
\end{align}
As in previous case, in order to ensure a well behaved spacetime at infinity as well as
having an event horizon, the dynamical exponent must belong to the following range
\[
z\in[1, \frac{3+\sqrt{5}}{2}[\, .
\]


\subsection{$\chi=\frac{1}{2}(z+1)$}

The last class of solutions is given for $\chi=\frac{1}{2}(z+1)$ with
\begin{align}
 \xi&=\frac{3 z^{2}-4 z+3}{2 \left( 9 z^{2}-12 z+11\right) }\, ,\nonumber\\
 \alpha&=\frac{c_1(z-3) ( 9z^2 -12 z-11)}{2 (z-1) (z^2-3 z+1)} \, ,\nonumber\\
  M&=\frac{(z-1) (21z^2-13z^2+31z -15) }{16 (9z^2 -12 z-11)}\, \nonumber\\
 \lambda&=-\frac{3(z-1)^{3} (z^2-3 z+1) (9z^2 -12 z-19)}{4 (z-3) (9z^2 -12 z-11)^{2}}\, .\nonumber
\end{align}
The dynamical exponent falls within the following range
\[
z\in[0, \frac{3-\sqrt{5}}{2}[\cup]1,\frac{3+\sqrt{5}}{2}[ \cup ]3, \infty[\,.
\]
In contrast with the previous cases, here there is wider range of possible values for the dynamical exponent. Note that none of these three classes of solutions accomodate
a stealth with $z=3$ \cite{AyonBeato:2009nh} as it was the case for the asymptotically AdS solution. As a last comment we also notice that minimally coupled scalar fields are not allowed
in the Lifshitz case.


\section{Conclusions}

In this note, we have reported two classes of black hole solutions of new
massive gravity in three dimensions with a source described in term of a
self-interacting and massive scalar field (non)minimally coupled. The first
metric family corresponds to an asymptotically AdS black hole in a specific
region of the coupling constants which are all parameterized in term of $\xi$.
As in the free source case, we have shown that the same equations also admit
Lifshitz black hole solutions for a different set of the parameters expressed
in this case in term of the dynamical exponent $z$. It is somehow appealing
that the equations of new massive gravity with or without source may
accommodate such classes of black holes with different asymptotic behavior. It is also
surprising that the matter source that has made possible the construction of
these black hole solutions is quite simple. Indeed, it involves a scalar
field $\Phi$ that can be massive, (non)minimally coupled and the
self-interacting potential is a physical one $U\propto\Phi^{4}$. We have also
computed the Wald formula for the entropy and realized that it is
proportional to the lapse metric function evaluated at the horizon. As a
consequence, the Wald entropy vanishes identically in spite of the fact that
the solutions have a non zero temperature. We have not computed the mass since
imposing that the first law of thermodynamics holds, this would yield
to a zero mass. Indeed, in Ref. \cite{Correa:2013bza}, there were obtained planar
black hole solutions for Lovelock gravity with a scalar field nonminimally
coupled which as well have zero entropy when computed using Wald's formula. In this case, using the Euclidean formalism,
it was explicitly shown that the mass indeed vanishes. Independently, it
will be nice to provide a complete thermodynamics analysis of the solutions
here derived. Also it will be desirable to understand the physical meaning of
these solutions which have a zero entropy. It seems that it is due to the
presence of higher-order curvature terms as well as to the fact that the transverse section of these solutions is planar. The scalar field found here is static
in both families of solutions, hence a rotating version of the AdS black hole
solution can easily be obtained by operating with an improper boost in the
$(t-\varphi)-$plane. This trick will also work in the case of the Lifshitz
black hole with the difference that the spinning version of the Lifshitz black
hole will violate the Lifshitz isometry at infinity. In three dimensions,
solitons can easily be constructed from static black holes by operating a
double Wick rotation. In the case of Lifshitz black holes with dynamical
exponent $z$, the corresponding soliton will enjoy the Lifshitz anisotropy
asymptotically with a dynamical exponent $z^{-1}$. These solitons may be
useful to better understand the thermodynamics issue of the solutions
presented here. Indeed, in Ref. \cite{Gonzalez:2011nz}, the authors proposed a
 generalization of the Cardy formula in order to compute the
semiclassical entropy of Lifshitz black hole with dynamical exponent $z$, and
in this formula, the ground state is played by the soliton with dynamical
exponent ${z^{-1}}$. \newline


\begin{acknowledgments}

FC is partially supported by
grant 11121651 from FONDECYT and from CONICYT 79112034.
MH is partially supported by
grant 1130423 from FONDECYT and from CONICYT,
Departamento de Relaciones Internacionales ``Programa Regional MATHAMSUD 13 MATH-05''. MH also acknowledge the CECs for his kind hospitality where part of this work has been done.
JO is partially supported by
grant 1141073 from FONDECYT. 
The Centro de Estudios Cient\'{\i}ficos (CECs) is funded by the Chilean government
through the Centers of Excellence Base Financing Program of Conicyt.

\end{acknowledgments}



\end{document}